\documentclass[nofootinbib,superscriptaddress,twocolumn,showpacs]{revtex4-2}
\usepackage{amssymb,amsmath,amsthm,graphicx,subfigure,float,cancel,braket,comment,tipa,bm,dsfont,graphicx,hyperref}
\usepackage{xcolor}
\usepackage{mathrsfs,enumitem} 
\usepackage{newtxtext,newtxmath}%
\hypersetup{
	colorlinks,
	linkcolor={red!50!black},
	citecolor={blue!50!black},
	urlcolor={blue!50!black}
}

\newcommand{\at}{\dot{a}}
\newcommand{\att}{\ddot{a}}
\newcommand{\Nt}{\dot{N}}

\newcommand{\p}{\partial}

\begin{document}
\title{\textbf{Quantum Einsteinian Cubic Cosmology}}
\author{Nephtalí Eliceo Martínez Pérez \thanks{nephtali.martinezper@alumno.buap.mx}}
\author{Cupatitzio Ramírez Romero \thanks{cramirez@fcfm.buap.mx}}
\affiliation{Facultad de Ciencias Físico Matemáticas, Benemérita Universidad Autónoma de Puebla\\
Puebla, México}

\begin{abstract}
    We study Cosmological Einsteinian Cubic Gravity (CECG) \cite{ARCINIEGA2020135272} in the context of minisuperspace quantum cosmology. CECG is a modification of Einstein's gravity by cubic curvature terms that yield a nontrivial contribution to the dynamics of FRW backgrounds while keeping the Friedmann equations at second order. First, we study the Hamiltonian formulation of the effective one-dimensional FRW CECG action using Ostrogradski's canonical variables and Dirac's algorithm for constrained systems. Since the momentum $p_a$ conjugate to the scale factor is a polynomial of degree five in $\at$, we implement canonical transformations $(a,p_a)\to (A,P)$ that enable us to write the Hamiltonian constraint explicitly. Second, we perform the Wheeler-DeWitt quantization using the new canonical variables. Although FRW CECG has no extra degree of freedom besides the scale factor, its non-standard Hamiltonian yields a higher-derivative Wheeler-DeWitt equation. We obtain exact solutions for the spatially flat case, and WKB-type solutions for the spatially closed case. Finally, we consider a homogeneous scalar field $\phi$ with inflationary potential and obtain WKB wave functions leading to strong correlations between coordinates and momenta.
\end{abstract}

\maketitle
\section{Introduction}
Higher-curvature modifications of Einstein's gravity have attracted a lot of attention in cosmology for they allow us to address long-standing problems such as the current accelerated expansion of the universe and the presumed inflationary phase that preceded the Hot Big Bang era \cite{NOJIRI20171,RevModPhys.82.451,defelice2010}. On a more fundamental level, it was early recognized the necessity of higher-curvature invariants for the renormalization of perturbative quantum gravity \cite{Stelle:1976gc}. From the perspective of effective field theory, Einstein's gravity is the leading term in the low-energy expansion of a yet to be seen quantum theory of gravity. The effective action includes an infinite series of higher-curvature terms suppressed by powers of a high-energy scale \cite{Kuntz}.

Generic higher-curvature theories lead to higher-derivative field equations, which signal the presence of extra degrees of freedom, including Ostrogradsky ghosts \cite{stelle78,Woodard2007}. However, there are theories expressly designed to satisfy second-order field equations such as Lovelock theories \cite{PADMANABHAN2013115}, which, in dimension $D$, introduce curvature densities of up to  order $\lfloor D/2 \rfloor$. For $D=2n$, however, the highest order term is topological, thus Lovelock yields the same dynamics as GR in four dimensions. 

A more relaxed condition is field equations being second-order only when evaluated on highly symmetric backgrounds such as static spherically symmetric ones, this is called Quasi Topological Gravity (QTG) and can be defined in $D\ge 5$ \cite{PhysRevD.82.124030, Oliva_2010,Myers2010}. Einsteinian Cubic Gravity (ECG) \cite{PhysRevD.94.104005}, on the other hand, is a theory in $D$-dimensions with up to cubic curvature terms satisfying the following requirements: (a) on maximally symmetric backgrounds, has the same linearized spectrum as general relativity, (b) the coupling constants are dimension-independent and (c) the extra cubic terms are neither trivial nor topological in $D=4$. ECG introduces a cubic curvature density $\mathcal{P}$ given below. Soon after, Generalized Quasi-Topological Gravity (GQTG) was developed \cite{Hennigar,Bueno2019,Bueno_2020}, which encompasses D-dimensional theories with $n$-order curvature terms admitting static spherically symmetric vacuum solutions characterized by a single field equation. Two other independent cubic densities $\mathcal{C}$ and $\mathcal{C}'$ were identified that contribute non-trivially in $D>4$.

For cosmological FRW backgrounds, the $\mathcal{P}$ and $\mathcal{C}$ cubic curvature densities yield separately fourth-order equations of motions. However, it was noticed that a particular combination of these terms yields second-order modified Friedman equations. This is called Cosmological Einsteinian Cubic Gravity (CECG) \cite{ARCINIEGA2020135272}. See also \cite{CISTERNA2020135435} for the quartic and quintic generalization, and \cite{Dengiz2025} for the conformal generalization. 
The cubic correction of CECG has the remarkable feature of generating accelerated expansion in the presence of ordinary matter. Although the expansion was of power-law type $a(t)\sim t^{3/2}$ (driven by radiation), exponential inflationary models were soon after constructed by introducing an infinite tower of curvature terms \cite{ARCINIEGA2020135242, arciniega2025geometriccosmology}. This approach is known as geometric inflation. (See also the $f(P)$-theory \cite{PhysRevD.99.123527, Marciu2022}).

On the other hand, Quantum Cosmology explores the quantum origin of the universe and its transition into an essentially classical system through decoherence mechanisms \cite{wiltshire2003introductionquantumcosmology,halliwell,moniz}. It addresses fundamental questions such as explaining the initial state of the universe. For instance, with single-field inflationary models there is a threshold for the initial inflaton value above which the number of e-folds generated is sufficient to solve the problems of standard cosmology \cite{baumann2012tasilecturesinflation}. Therefore, it is pertinent to ask how probable it was for the universe to start its evolution with a field value  favoring sufficient inflation. The wave function $\Psi(a,\phi)$ can be used to construct a probability distribution of initial values \cite{halliwell, mijic, higherderivatives}. 

Quantum cosmology is worked out in a quantum gravity framework. We work here in the context of Quantum Geometrodynamics, which is based on the ADM formulation of gravity \cite{bojowald}. The Hamiltonian has the form $H=\int d^3 x (N \mathcal{H}_0+N^i \mathcal{H}_i)$, where $\mathcal{H}_0\approx 0 \approx \mathcal{H}_i$ are the Hamiltonian and momentum constraints, respectively. Following Dirac's quantization approach, physical states must be annihilated by the quantum constraint operators. Defining the wave function $\Psi(h_{ij}(x))$ on superspace, that is, the set of spatial metrics $h_{ij}(x)$ modulo spatial diffeomorphisms, the remaining equation is $\mathcal{H}_0 \Psi(h_{ij})=0$, called Wheeler-DeWitt (WDW) equation. 

Since the full superspace WDW equation is extremely complicated, finite dimensional ``minisuperspace" WDW equations, resulting from the imposition of symmetries (prior to quantization) are usually considered. At the largest scales, the universe is well described by spatially homogeneous and isotropic or FRW models \cite{ellis}. As inhomogeneities tend to grow with time due to gravitational instability, going backwards in time, the universe only gets even more symmetric. In fact, a triumph of the inflationary paradigm is providing a mechanism to evolve quantum fluctuations into the late-time perturbation spectrum in energy and curvature densities \cite{baumann2012tasilecturesinflation}. Models with a certain degree of anisotropy, called Bianchi, are also frequently considered in quantum cosmology. These truncated models are still useful to explore technical and conceptual aspects of the different quantization approaches while offering insights into the quantum nature of the universe \cite{Kiefer:2004xyv}.

In this work, we consider the FRW minisuperspace model provided by CECG. As we will review, in absence of matter fields, only a positive cosmological constant, the effect of cubic curvature is encoded in a rescaled Hubble constant. In view of this, we ask whether the higher-derivative terms  complicate, or not, the quantization procedure and the outcomes. In particular, how different the wave function is compared to that of the ordinary (linear curvature) FRW model. As we shall see, the cubic curvature terms give rise to some interesting challenges already at the Hamiltonian level, whereas in the quantum scenario we face complicated non-polynomial operators. Nonetheless, we manage to obtain wave functions resembling those of the ordinary case. 

The paper is organized as follows. In Section \ref{sec2}, we obtain an effective 1D Lagrangian $L^{\text{CECG}}$ corresponding to FRW Cosmological Einsteinian Cubic Gravity. Since $L^{\text{CECG}}$ is linear in the acceleration $\att$, we integrate by parts to identify an ordinary Lagrangian $L(N,a,\at)$ leading to the same dynamics. The latter has a non-standard kinetic term depending on $\at^6$. In Section \ref{sec3}, we describe the Hamiltonian formulation of both lagrangians; for $L^{\text{CECG}}$ we use  Ostrogradski's canonical variables and Dirac's algorithm, which yields a non-trivial Dirac's bracket between $a$ and the momentum of the ordinary theory, $p\sim a \at$. Now, if we use the alternative Lagrangian, we are confronted immediately with a conjugate momentum $p_a$ given by a polynomial of degree five in $\at$. Since there is no general solution to the quintic equation (with a finite number of algebraic operations \cite{Linton}), we resort to alternative phase space variables $(A,P)$ arising from canonical transformations. In this way, we trade the problem of solving for $\at(a,p_a)$ for the problem of finding the inverse canonical transformation. In Section \ref{sec4}, we quantize canonically the FRW CECG Hamiltonian expressed in terms of the new canonical pairs. We obtain simple analytic solutions of the WDW equation for the spatially flat case. There are exponential solutions with imaginary phase which can be associated with the classical solutions of the Friedmann equations, as well as complex exponential solutions, absent in the ordinary FRW case with no classical counterpart. In Section \ref{sec5}, we obtain WKB type wave functions for the spatially closed model using another pair $(X,\Pi)$, which as the usual FRW, are characterized by a scale $\bar{X}$, where the wave function changes from oscillatory to exponential behavior. In Section \ref{sec6}, we introduce a scalar field with Starobinsky potential, describe classical inflationary solutions  and obtain WKB wave functions exhibiting exponential and oscillatory behavior. Finally, in Section \ref{sec7} we draw some conclusions and comment on interesting aspects that have been left out in this paper.

\section{FRW action}\label{sec2}
The action of Cosmological Einsteinian Cubic Gravity with only a positive cosmological constant $\Lambda$ is \cite{ARCINIEGA2020135272}
\begin{align}\label{ecg}
	S^{\text{CECG}}=\int d^4x\, \sqrt{-g} \left[\frac{R-2 \Lambda}{2 \kappa^2}+\beta (\mathcal{P}-8\, \mathcal{C})\right],
\end{align}
where $\kappa^2=8\pi G=8 \pi \, m_P^{-2}$ (in natural units $c=1=\hbar$), $\beta$ is a constant of dimension length squared, and the cubic densities are \cite{PhysRevD.94.104005, Hennigar}
\begin{align}
	\mathcal{P}&=12 R_{a\ b}^{\ c\ d} R_{c\ d}^{\ e\ f} R_{e\ f}^{\ a \ b}+R_{ab}^{\ \ cd} R_{cd}^{\ \ ef} R_{ef}^{\ \ ab} \nonumber \\
	&-12 R_{abcd} R^{ac} R^{bd}+8 R^b_{\ a} R^c_{\ b} R^a_{\ c}, \\
	\mathcal{C}&=R_{abcd} R^{abc}_{\ \ \ e} R^{de}-\frac{1}{4} R_{abcd} R^{abcd} R \nonumber \\
	&-2 R_{abcd} R^{ac} R^{bd}+\frac{1}{2} R_{ab} R^{ab} R.
\end{align}

For the minisuperspace model, we evaluate action (\ref{ecg}) with an FRW metric $ds^2=-N^2(t)+a^2(t) [dr^2/(1-k r^2)+r^2 (d\theta^2+\sin^2 \theta d\phi^2)]$. The lapse function $N(t)$ is the gauge field of time-reparametrizations whereas the scale factor $a(t)$ tracks the overall expansion of the spatial slices.  Next, integrating the spatial coordinates over a finite volume $\mathcal{V}$ and switching to a scale factor with length dimension $a\to a \mathcal{V}_0^{1/3}$ (and dimensionless $k$), we obtain the effective 1D CECG Lagrangian (cf. the related $f(P)$ model ($k=0$) \cite{PhysRevD.99.123527})
\begin{align}\label{ecgfrw}
	L^{\text{CECG}}&=\frac{3}{\kappa^2} N a^3 \left( \frac{\dot{B}}{a N}+\frac{B^2+k}{a^2}-\frac{\Lambda}{3}\right)  \nonumber \\
	&\quad +48 \beta  N \frac{\left(B^2+k\right)^2}{a^3} \left( B^2+k-3 \frac{a \dot{B}}{N}\right) ,
\end{align}
using the notation $B\equiv \frac{\at}{N}$. 

The CECG action possesses symmetric criticality \cite{Palais1979, Deser2003} with respect to the FRW isometries, that is, evaluating the set of field equations on an FRW metric reduce it to a subset of non-trivial equations that can also be derived from the 1D FRW action (\ref{ecgfrw}). 

Now, as it stands, Lagrangian (\ref{ecgfrw}) depends on $\att$ and dynamics is thus governed by higher-derivative Lagrange equations \cite{Sundermeyer2014}. Nonetheless, since the acceleration only appears linearly, the equation of motion for $a$ is actually second-order. To see this, we integrate by parts (\ref{ecgfrw}) in the following schematic way (recall $B=\at/N$)
\begin{align}\label{decompose}
	L^{\text{CECG}}&=N G(a, B)+\dot{B} \frac{\p F(a,B)}{\p B} \nonumber \\
	&=N \left[ G(a,B)-B \frac{\p F(a,B)}{\p a}\right]+\frac{dF(a,B)}{dt} \nonumber \\
	&\equiv L(N, a,B)+\frac{dF(a,B)}{dt},
\end{align}

Comparing the first equality of (\ref{decompose}) with (\ref{ecgfrw}), one readily identifies $\frac{\p F}{\p B}$, whose integration, up to an additive function of $a$, yields
\begin{align}\label{explain}
	F(a,B)=\frac{3 a^2 B}{\kappa^2}-\frac{144 \beta}{a^2} \left(\frac{B^5}{5}+\frac{2}{3} k B^3+k^2 B \right).
\end{align}

The time derivative of (\ref{explain}) does not contribute to the (higher-derivative) Lagrange equations but does affect the boundary terms arising upon variation of the action. For instance, in the gauge $N=1$, 
\begin{align}\label{boundary}
	\delta F(a, \at)|_{t_i}^{t_f}=\left. \left(\delta a \frac{\p F}{\p a}+\delta \at \frac{\p F}{\p \at} \right) \right|_{t_i}^{t_f} 
\end{align}
Thus, to get rid of (\ref{boundary}) requires fixing $a$ and $\at$ at the initial and final times.  

To avoid higher-derivative Lagrange equations and their extra boundary conditions, while ensuring functional differentiability, it is standard practice to redefine the action with suitable boundary terms, e.g., the Gibbons-Hawking-York term for the Einstein-Hilbert action \cite{bojowald}. In our case, this amounts to drop the total derivative and work with the alternative Lagrangian defined in (\ref{decompose}), namely,  
\begin{align}\label{alter}
	L&=\frac{3}{\kappa^2} N \left(-a B^2+k a-\frac{\Lambda}{3} a^3\right) \nonumber \\
	&\quad -48 \beta N \frac{\left(\frac{1}{5} B^6+B^4 k+3 B^2 k^2-k^3 \right)}{a^3}.
\end{align}

In the proper time gauge $N=1$, (\ref{alter}) yields the second order equation of motion \cite{ARCINIEGA2020135272}
\begin{align}\label{acceleration}
\frac{\att}{a}=\frac{\Lambda-a^{-2} \big(\at^2+k\big)+48 \beta \kappa^2 a^{-6} \big(\at^2+k\big)^3}{2+96 \beta \kappa^2 a^{-4} \big(\at^2+k\big)^2}. 
\end{align}
provided that the Hessian condition holds,  
\begin{align}\label{hess}
	\frac{\p^2 L}{\p \at^2}=-\frac{6 a}{\kappa^2} \left[1+48 \beta \kappa^2 a^{-4} \big(\at^2+k \big)^2\right]  \ne 0.
\end{align}

Since (\ref{alter}) does not depend on $\Nt$, we also get a constraint, $\frac{\p L}{\p N}=0$, which yields the modified Friedmann equation, 
\begin{align}\label{friedcanon}
	\frac{\at^2}{a^2}+\frac{k}{a^2}+16 \beta \kappa^2 \left(\frac{\at^2}{a^2}+\frac{k}{a^2}\right)^3=\frac{\Lambda}{3}.
\end{align}

Substituting de Sitter type solutions (see \cite{z4z2-d281} for the maximally symmetric solutions of ECG), 
\begin{align}\label{classicsol}
	a(t)=
	\begin{cases}
		\exp (\alpha t), \quad k=0, \\
		\alpha^{-1} \cosh (\alpha t), \quad k=1 \\
		\alpha^{-1} \sinh (\alpha t), \quad k=-1
	\end{cases}
\end{align}
into (\ref{friedcanon}) yields the algebraic equation 
\begin{align}\label{cubicw}
	\alpha^2+16 \beta \kappa^2 \alpha^6=\frac{\Lambda}{3}.
\end{align}

\begin{figure}[h!]
	\centering
	\includegraphics[scale=0.6]{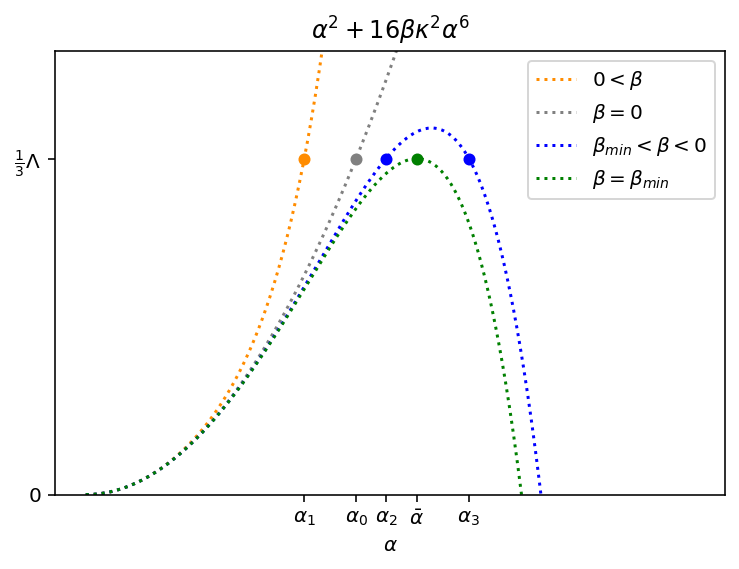}
	\caption{The left-hand side of (\ref{friedcanon}) for different ranges of $\beta$.}
	\label{gig0}
\end{figure}

The polynomial of $\alpha$ on the left-hand side of (\ref{cubicw}) is shown in Fig. \ref{gig0}. With $\beta=0$, $|\alpha|=|\alpha_0|=\sqrt{\Lambda/3}$. With negative $\beta$, the curve is bounded from above and (\ref{cubicw}) has real solutions only if \cite{PhysRevD.94.104005}
\begin{align}\label{realsol}
\beta \ge -\frac{1}{12 \kappa^2 \Lambda^2}=\beta_{\text{min}}
\end{align}
At $\beta=\beta_{\text{min}}$, $|\alpha|=|\bar{\alpha}|=\sqrt{\Lambda/2}=(-48 \beta_{\text{min}} \kappa^2)^{-1/4}$ (in this case the Hessian (\ref{hess}) vanishes).

On the other hand, using the general solution of the cubic equation \cite{cubic}, the real roots ($\alpha^2>0$) of (\ref{cubicw}), are given by 
\begin{subequations}\label{roots}
\begin{align}
	\alpha_1^2&=\frac{\eta^{1/3}-\eta^{-1/3}}{\sqrt{48 \beta \kappa^2}}, && \beta>0, \label{alphaplus}	\\
	\alpha^2_{2,3}&=\frac{\cos \big[\frac{1}{3} \tan^{-1} \sqrt{\frac{1+Q}{-Q}} \pm \frac{\pi }{3}\big]}{\sqrt{-12\beta \kappa^2}}, && \beta<0, \label{alphaminus}
\end{align}
\end{subequations}
where we use the following definitions,
\begin{align}\label{defs}
	\begin{split}
		& Q=12 \beta \kappa^2 \Lambda^2, \\
		& \eta=\sqrt{1+Q}+\sqrt{Q},
	\end{split}
\end{align}
and $\alpha_2, \alpha_3$  correspond to $+, -$ signs in (\ref{alphaminus}), respectively. 

\begin{figure}[h!]
	\centering
	\includegraphics[scale=0.6]{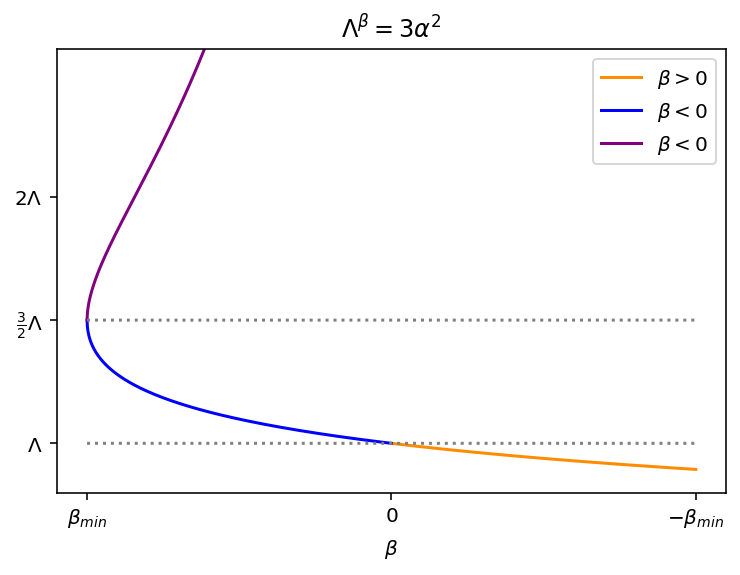}
	\caption{Plot of $\Lambda^\beta(\beta)$ according to (\ref{effla}). Blue/purple segments correspond to $\alpha_2/\alpha_3$ in (\ref{alphaminus}), respectively.}
	\label{gig6}
\end{figure}

Given that $\Lambda=3 \alpha^2$ for $\beta=0$, one can define an effective cosmological constant depending on $\beta$ (cf. \cite{z4z2-d281}),
\begin{align}\label{effla}
	\Lambda^\beta \equiv 3 \alpha^2(\beta),
\end{align}
which is depicted in Fig. \ref{gig6}. Thus $\Lambda^\beta$ is smaller than $\Lambda$ for positive $\beta$ and vice versa. Note the two branches for negative $\beta$. 

Finally, for $\Lambda=0$ and negative $\beta$, besides the flat spacetime solution ($\alpha=0$), there are solutions of the form (\ref{classicsol}) with an effective cosmological constant induced completely by the cubic curvature terms,
\begin{align}\label{vaccum}
	\alpha^2=\frac{1}{\sqrt{-16 \beta \kappa^2}}.
\end{align}

\section{Hamiltonian formulation}\label{sec3}
Since we are interested in Wheeler-DeWitt quantization, the Hamiltonian formulation of FRW CECG is in order. First, we consider Lagrangian (\ref{ecgfrw}). Since it depends on $\att$, we implement Ostrogradski's Hamiltonian construction \cite{Woodard2007}, and then perform Dirac's algorithm for constrained systems \cite{henneaux}. We start with six canonical variables
\begin{align}\label{ostrovar}
\begin{split}
a, \qquad p_a&\equiv \frac{\p L^{\text{CECG}}}{\p \at}-\frac{d}{dt} \frac{\p L^{\text{CECG}}}{\p \att}\\
v \equiv \at, \qquad p_v&\equiv \frac{\p L^{\text{CECG}}}{\p \att}, \\
N, \qquad p_N&=\frac{\p L^{\text{CECG}}}{\p \dot{N}}.
\end{split}
\end{align}

The special form of the Lagrangian yields three (primary) constraints $C_i(N,a, v, p_N, p_a, p_v)\approx 0$ ($i=1,2,3$) where
\begin{align}\label{list}
	\begin{split}
		C_1&=p_N-\frac{144 \beta v}{a^2 N^2} \left(\frac{v^2}{N^2}+k \right)^2, \\
		C_2&=p_a+\frac{6 a v}{\kappa^2 N}, \\
		C_3&=p_v+\frac{144 \beta}{a^2 N} \left(\frac{v^2}{N^2}+k \right)^2. 
	\end{split}
\end{align}

The Ostrogradski Hamiltonian is defined as follows
\begin{align}\label{hamil}
H&\left(N,a,\frac{v}{N}\right) =\at p_a+\att p_v+\dot{N} p_N-L^{\text{ECG}}  \nonumber \\
&=N a^3 \left[\frac{\Lambda}{\kappa^2}-\frac{3}{\kappa^2} \frac{\big(\frac{v^2}{N^2}+k\big)}{a^2}-48 \beta \frac{\big(\frac{v^2}{N^2}+k\big)^3}{a^6}\right]
\end{align}

The total Hamiltonian is $H^{\text{tot}}=H+u_1 C_1+u_2 C_2+u_3 C_3$ with so far undetermined multipliers $u_i$. Next, we impose consistency conditions $\dot{C}_i =\{C_i, H^{\text{tot}}\} \approx 0$ on the primary and any other arising constraint. The result of this process is that two of the multipliers get determined, $u_2=v$, $u_3=\frac{v}{N} u_1+\frac{v^2}{a}+N^2 \frac{k}{a}$, $u_1$ remains arbitrary. Further, we get a fourth (secondary) constraint $C_4\approx 0$, where  
\begin{align}\label{c4}
	C_4=-\frac{H}{N}.
\end{align}

Next, we arrange the constraints into first, $\phi_i$, and second class, $\chi_i$, as follows
\begin{align}
\begin{split}
	&\phi_1=C_1+\frac{v}{N} C_3=\frac{v p_v}{N}+p_N,\\
    &\phi_2=-C_4+\frac{v}{N} C_2+\left(\frac{v^2}{a N}+N \frac{k}{a}\right)  C_3,\\
    &\chi_1=C_2,\\
    &\chi_2=N C_3.
\end{split}
\end{align}
The Poisson bracket of the second-class constraints is
\begin{align}\label{second}
	\left\lbrace \chi_1,\chi_2\right\rbrace_P=\frac{6 a}{\kappa^2} \left[1+48 \beta \kappa^2 \left(\frac{v^2}{a^2 N^2}+\frac{k}{a^2}\right)^2\right],
\end{align}
which was assumed be to nonzero (see (\ref{hess})).

Therefore, the total Hamiltonian is a combination of first class-constraints $H^{\text{tot}}=N \phi_2+u_1 \phi_1$, 
and the counting of degrees of freedom gives $n=3-2-\frac{2}{2}=0$, as with the linear FRW model.

Using Dirac's brackets, we can set the second-class constraints strongly equal to zero, $\chi=0$. Defining $\tilde{p}_N\equiv \phi_1|_{\chi=0}$ and $H_0\equiv \phi_2|_{\chi=0}$, the  total Hamiltonian acquires the standard form $H^{\text{tot}}=N H_0+u_1 \tilde{p}_N$, where 
\begin{align}\label{hamihhigher}
	H_0=a^3 \left[\frac{\Lambda}{\kappa^2}-\frac{3}{\kappa^2} \frac{\left(B^2+k\right)}{a^2}-48 \beta \frac{\left(B^2+k\right)^3}{a^6} \right]
\end{align}
and $B=\frac{\at}{N}=\frac{v}{N}$. The basic Dirac's brackets are  
\begin{align}
&\{N,\tilde{p}_N\}_D=1, \\
&\left\lbrace a, B \right\rbrace_D=-\frac{\kappa ^2}{6 a \big[1+48 \beta \kappa^2 a^{-4} (B^2+k)^2\big] }. \label{aB}
 \end{align}
The Hamiltonian equations of motion $\dot{a}=\{a, H^{\text{tot}}\}_D$ and $\dot{B}=\{B, H^{\text{tot}}\}_D$, reproduce equation (\ref{acceleration}).

Now, let's consider the Hamiltonian formulation of Lagrangian (\ref{alter}). Since there is no dependence on the acceleration, we proceed with the usual definition of conjugate momenta. The four canonical variables are 
\begin{align}\label{alterpa}
\begin{split}
&N, \quad p_N=0, \\
&a, \quad p_a=-\frac{6 a B}{\kappa^2}-\frac{48 \beta \big(\frac{6}{5} B^5+4 k B^3+6 k^2 B \big)}{a^3}. 
\end{split}
\end{align}

Here we face the problem that there is no general solution of the quintic equation \cite{Linton} to solve for $\at(a,p_a)$. Even if there was an algebraic solution, it would yield a branched Hamiltonian \cite{Teitelboim_1987, Bagchi_2024}. With the higher-derivative Lagrangian this difficulty is avoided since $\at$ is treated as an independent canonical variable.

The usual Hamiltonian is 
\begin{align}\label{hamilvel}
	\tilde{H}(N,a,p_a)&\equiv \at p_a+\dot{N} p_N-L \nonumber \\
	&=\left. H(N,a,B)\right|_{B=B(a,p_a)}
\end{align} 
where $H$ is the function of three variables defined in (\ref{hamil}). In this approach, we only have first-class constraints, $p_N\approx 0$ and 
\begin{align}
	\dot{p}_N=-\frac{\p \tilde{H}(a,N,p_a)}{\p N}=-\frac{\p H(N,a,B)}{\p N}=-H_0 \approx 0.
\end{align}
with $H_0(A,B)$ defined in (\ref{hamihhigher}).

For an explicit expression of the Hamiltonian, we opt for performing a canonical transformation. For instance, 
\begin{align}\label{AP}
	\begin{split}
		A&=a^3 \Big[1+48 \beta \kappa^2 \frac{\left(B^4+6 k B^2-3 k^2 \right)}{a^4}\Big], \\
		P&=-\frac{2 B}{\kappa^2 a}.
	\end{split}
\end{align}
That (\ref{AP}) is a canonical transformation can be shown in several ways \cite{torres}. For example,  using the auxiliary set of (general) phase space coordinates $(\text{a}=a,B=\at/N)$, the Poisson bracket between $A$ and $P$ is   
\begin{align}\label{check}
	\{A,P\}=\begin{vmatrix}
		\frac{\p A}{\p a} & \frac{\p A}{\p p_a} \\[3pt]
		\frac{\p P}{\p a} & \frac{\p P}{\p p_a}
	\end{vmatrix} = \begin{vmatrix}
	\frac{\p A}{\p \text{a}} & \frac{\p A}{\p B} \\[3pt]
	\frac{\p P}{\p \text{a}} & \frac{\p P}{\p B}
	\end{vmatrix} \{\text{a},B\}=1,
\end{align} 
where we use the Poisson analogue of (\ref{aB}), namely, 
\begin{align}\label{nonhamil}
	\{\text{a},B\}&=\begin{vmatrix}
		\frac{\p \text{a}}{\p a} & \frac{\p \text{a}}{\p p_a} \\[3pt]
		\frac{\p B}{\p a} & \frac{\p B}{\p p_a}
	\end{vmatrix}=\begin{vmatrix}
		\frac{\p a}{\p \text{a}} & \frac{\p a}{\p B} \\[3pt]
		\frac{\p p_a}{\p \text{a}} & \frac{\p p_a}{\p B}
	\end{vmatrix}^{-1} \nonumber \\
	&=\frac{-\kappa^2}{6 \text{a} \left[ 1+48 \beta \kappa^2 \text{a}^{-4} \left(B^2+k\right)^2\right] }, 
\end{align}

Now, since transformation (\ref{AP}) does not involve time, we may choose (\ref{hamilvel}), expressed in terms of variables (\ref{AP}), as the new Hamiltonian. 

For $k=0$, combining (\ref{alterpa}) and (\ref{AP}), the inverse canonical transformation is given by
\begin{align}\label{new}
	\begin{split}
		a&=\frac{A^{1/3}}{(1+3 \beta \kappa^{10} P^4)^{1/3}}, \\
		p_a&=\frac{3 A^{2/3}}{(1+3 \beta \kappa^{10} P^4)^{2/3}} \left(P+\frac{3}{5} \beta \kappa^{10} P^5 \right).
	\end{split}
\end{align}

Therefore, the Hamiltonian constraint (\ref{hamihhigher}) reads
\begin{align}\label{hamil2}
	H_0&=\frac{-\frac{3}{4} \kappa^2 A P^2 \left(1+\beta \kappa^{10} P^4\right)+\frac{\Lambda}{\kappa^2} A}{1+3 \beta \kappa^{10} P^4} \approx 0,
\end{align}

\begin{figure}[h!]
	\centering
	\includegraphics[scale=0.54]{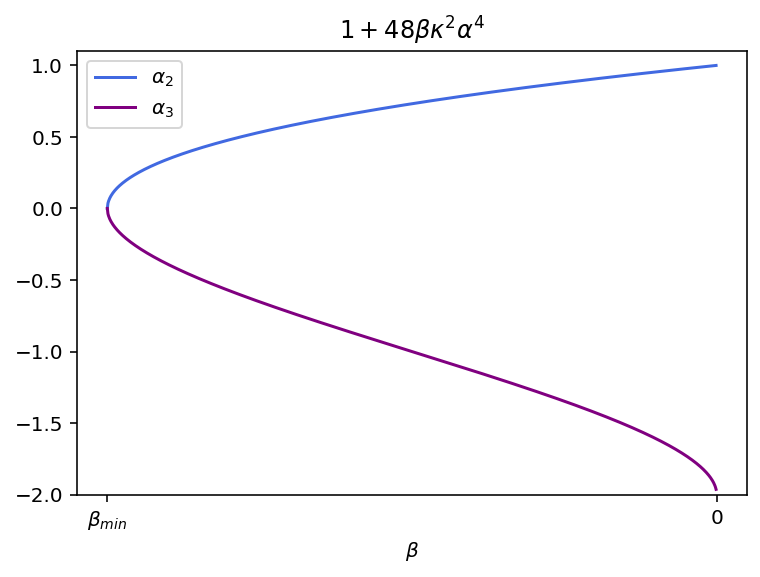}
	\caption{Hessian determinant (\ref{hess}) evaluated on the two different solutions (\ref{alphaminus}).}
	\label{gig5}
\end{figure}

The denominator in (\ref{new}) is proportional to the Hessian determinant (\ref{hess}) and for negative $\beta$ it can be either positive or negative. In fact, evaluating it on the classical solutions (\ref{alphaminus}), it ranges between 0 and -2,  as shown in Fig. \ref{gig5}. 

To conclude this section, it can be shown,  using the basic Dirac's bracket (\ref{aB}), that $\{a,p_a\}_D=1=\{A,P\}_D$, for the variables defined in (\ref{alterpa})-(\ref{AP}). Thus, on the reduced phase space  we get equivalent canonical formulations.

If we were to quantize directly (\ref{hamihhigher}), we would be prompted to find operators $\hat{a}, \hat{B}$ satisfying the commutator version of (\ref{aB}). We do not pursue that direction; instead, we will quantize the classical system expressed in terms of the alternative canonical variables $(A,P)$. Although the transformation (\ref{new}) is canonical, it is also highly non-linear, thus the existence of a unitary transformation relating $\tilde{H}_0(\hat{a},\hat{p}_a)$ to $H_0(\hat{A},\hat{P})$ is not guaranteed due to e.g., ordering ambiguities or the Groenewold–Van Hove obstruction \cite{10.1063/1.532854}.

\section{Wheeler-DeWitt equation}\label{sec4}
Let's consider the canonical quantization of the spatially flat model ($k=0$) in the ``position" representation: $\hat{P} \Psi(A)=-i \hbar \p_A \Psi(A)$ and $\hat{A} \Psi(A)=A \Psi(A)$. 

The actual form of the WDW equation depends on the ordering of the Hamiltonian $\hat{H}_0$. First, the term $D^{-1} \equiv (1+3 \beta \kappa^{10} P^4)^{-1}$ in (\ref{hamil2}) is promoted to a non-local operator related to integration. Let's begin defining $\widehat{D^{-1}}$  as a right-inverse of $\hat{D}\equiv \hat{I}+3 \beta \kappa^{10} \hbar^4 \p_A^4$, that is, 
\begin{align}\label{resolv}
\hat{D} \big[\widehat{D^{-1}} \psi(A)\big]=\psi(A). 
\end{align}
However, in reverse order we do not get the identity yet, rather $\widehat{D^{-1}} [\hat{D} \psi(A)]=\psi(A)+\chi(A)$, with $\chi(A) \in \text{Ker}(\hat{D})$.  For a true inverse, the domain of $\hat{D}$ must be restricted by means of boundary conditions, thus defining a $\hat{D}_\circ$ such that $\text{Ker}(\hat{D}_\circ)=\{0\}$  \cite{Teschl}. In that case, $\widehat{D^{-1}}$ can be defined as the inverse of $\hat{D}_\circ$. In particular, $\phi(A)\equiv \widehat{D^{-1}} \psi(A)$ can be expressed as 
\begin{align}
\phi(A)=\int_0^{\infty}	 G(A,A') \psi(A') dA',
\end{align}
in terms of a unique Green's function, i.e., $\hat{D} G(A,A')=\delta(A-A')$, satisfying the prescribed boundary conditions. 

For instance, for positive $\beta$ and $A\ne A'$, equation $\hat{D} G(A,A')=0$ admits  solutions of the form $e^{\pm \frac{A}{c}} \cos \frac{A}{c}$, $e^{\pm \frac{A}{c}} \sin \frac{A}{c}$, where $c=\sqrt[4]{12 \beta \kappa^{10}} \hbar$. Let $G^>, G^<$ denote $G(A,A')$ for $A>A'$ and $A<A'$, respectively, then, the boundary conditions, $G^<|_{A=0}=0=\p_A G^<|_{A=0}$, $G^>|_{A\to \infty}=0$, together with the matching rules at $A=A'$, namely, $G^>-G^<=0$, $\p_AG^>-\p_A G^<=0$, $\p_A^2 G^>-\p_A^2 G^<=0$ and $\p_A^3 G^>-\p_A^3 G^<=1/3 \beta \kappa^{10} \hbar^4$, we obtain
\begin{subequations}
\begin{align}
G^<(A,A')&=\frac{1}{2c} e^{-\frac{A+A'}{c}} \Big[-e^{\frac{2 A}{c}} \sin \frac{A-A'}{c}-\sin \frac{A+A'}{c} \nonumber \\
&+\left(e^{\frac{2 A}{c}}-2\right) \cos \frac{A-A'}{c}+\cos \frac{A+A'}{c}\Big],\\
G^>(A,A')&=\frac{1}{2c} e^{-\frac{A+A'}{c}} \Big[e^{\frac{2 A'}{c}} \sin \frac{A-A'}{c}-\sin \frac{A+A'}{c} \nonumber \\
&+\left(e^{\frac{2 A'}{c}}-2\right) \cos \frac{A-A'}{c}+\cos \frac{A+A'}{c}\Big].	
\end{align}
\end{subequations}

These considerations show that the generic WDW equation arising from (\ref{hamil2}) is an integro-differential equation. Nonetheless, with a convenient ordering we get to solve just an ordinary differential equation. For instance, writing the Hamiltonian constraint operator as  
\begin{align}\label{wdw}
	\widehat{D^{-1}} A \left(-\frac{3 \kappa^2}{4} \left(\hat{P}^2+\beta \kappa^{10} \hat{P}^6\right)+\frac{ \Lambda}{\kappa^2} \right) \Psi=0
\end{align}
Then, applying $\hat{D}$ on both sides of (\ref{wdw}) yields
\begin{align}\label{wdw2}
	A \left(\hat{P}^2+\beta \kappa^{10} \hat{P}^6-\frac{4 \Lambda}{3 \kappa^4}\right) \Psi=0,
\end{align}
which is a quantum version of the Friedmann equation (\ref{friedcanon}), $\widehat{a^{-3} H_0} \Psi=0$. Next, let $\Psi(A)$ be the general solution of (\ref{wdw2}), substituting it back on the left-hand side of (\ref{wdw}) yields $\widehat{D^{-1}} (0)$. In general, $\widehat{D^{-1}}(0)$ can be a nontrivial function in $\text{Ker}(\hat{D})$, however, assuming boundary conditions, such as those of the example above, have been imposed so that $\text{Ker}(\hat{D})=\{0\}$, then $\widehat{D^{-1}} (0)$ is the zero function and the WDW equation (\ref{wdw}) holds.

\begin{figure}[h!]
	\centering
	\includegraphics[scale=0.58]{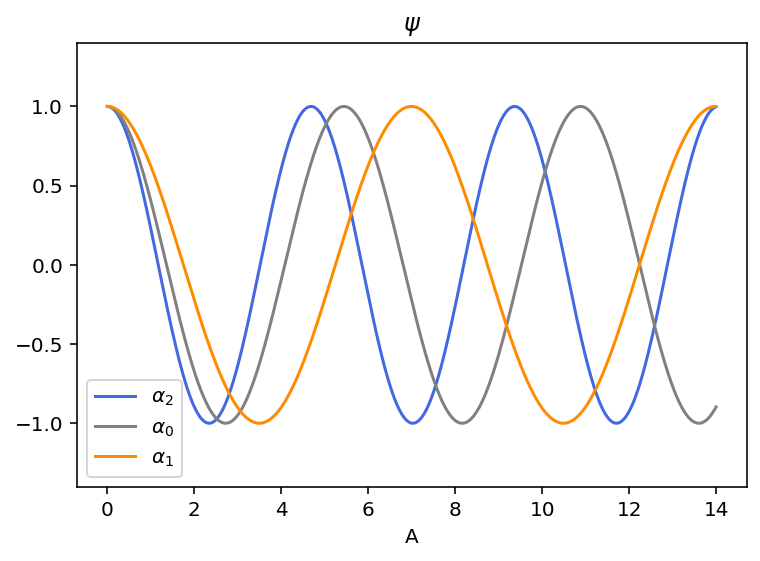}
	\caption{$\psi(A)=\cos(R A/\hbar)$ constructed with the real roots of (\ref{six}). With negative $\beta$, we restrict $A\in \mathbb{R}^+$ according to the sign of the factor $(1+48 \beta \kappa^2 H^4)$ in (\ref{AP}) (see Fig. \ref{gig5}).}
	\label{gig2}
\end{figure}

Now, (\ref{wdw2}) amounts to the following sixth-order differential equation
\begin{align}
	\left(\hbar^2\, \p_A^2+\beta \kappa^{10} \hbar^6\, \p_A^6+\frac{4 \Lambda}{3 \kappa^4} \right) \Psi(A)=0,
\end{align}
which admits simple solutions of the form 
\begin{align}\label{exp}
    \Psi(A)=\sum_{k=1}^6 c_k \exp \left(\frac{i}{\hbar} R_k A\right),
\end{align}
where parameters $R_k$ satisfying the algebraic equation
\begin{align}\label{six}
R^2+\beta \kappa^{10} R^6-\frac{4 \Lambda}{3 \kappa ^4}=0.
\end{align}
The real roots of (\ref{six}) are $R=2 \alpha/\kappa^2$, in terms of the $\alpha$'s in (\ref{roots}). Therefore, with (negative) positive $\beta$ the wavelength $\lambda=2\pi/R\propto \alpha^{-1}$ is  (smaller) larger than the ordinary FRW value $\lambda \propto \sqrt{3/\Lambda}$. See Fig. \ref{gig2}.

Of course, wave functions (\ref{exp}) are non-normalizable, for they represent an unrealistic situation of a universe free of matter fields. Nonetheless, we can make sense of these solutions through the semiclassical approximation \cite{halliwell}. In fact, because of our simple ordering choice (\ref{wdw2}), (\ref{exp}) coincides with the zeroth-order WKB approximation $\Psi_{\text{WKB} (0)}=\exp (\frac{i}{\hbar} S_0(A))$, where $S_0(A)=R A$ satisfies the Hamilton-Jacobi (H-J) equation corresponding to (\ref{hamil2}). In H-J formalism, $P=\frac{\p S_0}{\p A}$. Thus, identifying $\frac{\p S_0}{\p A}$ with the definition of the momentum (\ref{AP}) yields $ \frac{\at}{a N}=\pm \alpha R$, whose integration, in the gauge $N=1$ yields the $k=0$ solution in (\ref{classicsol}).

\begin{figure}[h!]
	\centering
	\includegraphics[scale=0.58]{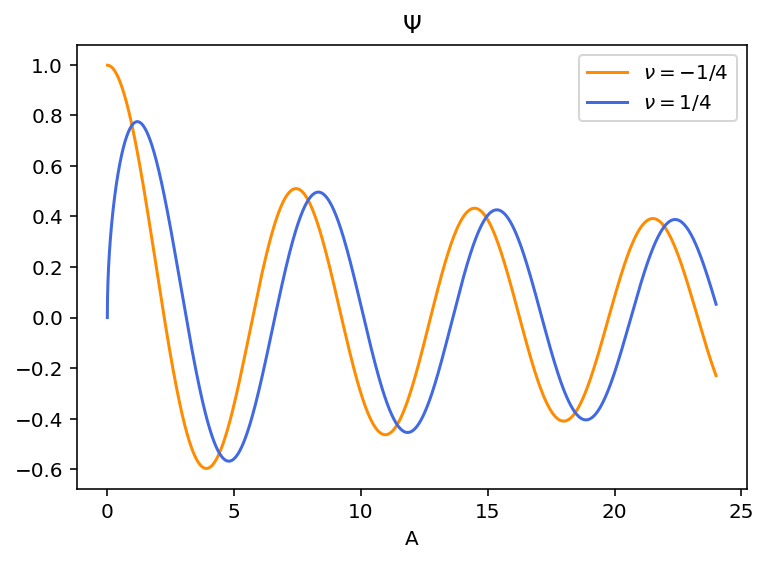}
	\caption{Exact solutions (\ref{analytic}) of the WDW equation with positive $\beta$. For large $A$, the amplitude decreases as $A^{-1/4}$.}
	\label{gig10}
\end{figure}

We may introduce a somewhat more sophisticated ordering of $P^2$ in (\ref{wdw2}) such as $A^{-n} P A^n P$. For example, if we choose $P^2\to A^{-1/2} \hat{P} A^{1/2} \hat{P}$, then (\ref{wdw2}) translates into
\begin{align}\label{lpcubic}
	\left[\p_A^2+\frac{1}{2 A} \p_A+\beta \kappa^{10} \hbar^4 \big(\p_A^2+\frac{1}{2 A} \p_A\big)^3+\frac{4 \Lambda}{3 \kappa^4 \hbar^2} \right] \Psi=0,
\end{align}
which admits solutions that are finite at $A=0$ given by Bessel functions of the first kind $J_{\pm \nu}(RA/\hbar)$,
\begin{align}\label{analytic}
	\Psi(A)=A^{1/4} \left[c_1 J_{-\frac{1}{4}} \left(\frac{R A}{\hbar} \right)+c_2 J_{\frac{1}{4}} \left(\frac{R A }{\hbar} \right)\right] ,
\end{align}
($c_1$ and $c_2$ are constants) depicted in Fig. \ref{gig10}. With this ordering, upon setting $\beta=0$, we retrieve the ordinary FRW wave functions with Laplace-Beltrami operator \cite{Kiefer:2004xyv}.

\subsection{Pure quantum states}
Now, we consider the complex solutions of (\ref{six}). For negative $\beta$, there are two purely imaginary roots $R=\pm i 2 \alpha/\kappa^2$ ($\alpha$ given by (\ref{alphaminus}) without the $\frac{1}{3} \pi$ phase) leading to exponentially growing/decaying wave functions. 

More interesting are the four complex roots of (\ref{six}) that exist for positive $\beta$. They can be expressed as $i R/\hbar=\pm \rho \exp(\pm i \theta)$, with the magnitude and phase given by 
\begin{align}\label{polar}
	\begin{split}
		\rho&=\frac{1}{\sqrt[4]{3 \beta \kappa^{10}} \hbar}\left[\eta^{-2/3}+1+\eta^{2/3}\right]^{1/4}, \\
		\theta&=\frac{1}{2} \tan^{-1} \left[\sqrt{3} \frac{\eta^{2/3}+1}{\eta^{2/3}-1}\right],
	\end{split}
\end{align}
using definitions (\ref{defs}). As $\beta$ runs over $(0,\infty)$, $\theta$ goes from $\frac{\pi}{4}$ to $\frac{\pi}{6}$, while $\rho$ goes from $\infty$ to $0$.

\begin{figure}[h!]
	\centering
	\includegraphics[scale=0.55]{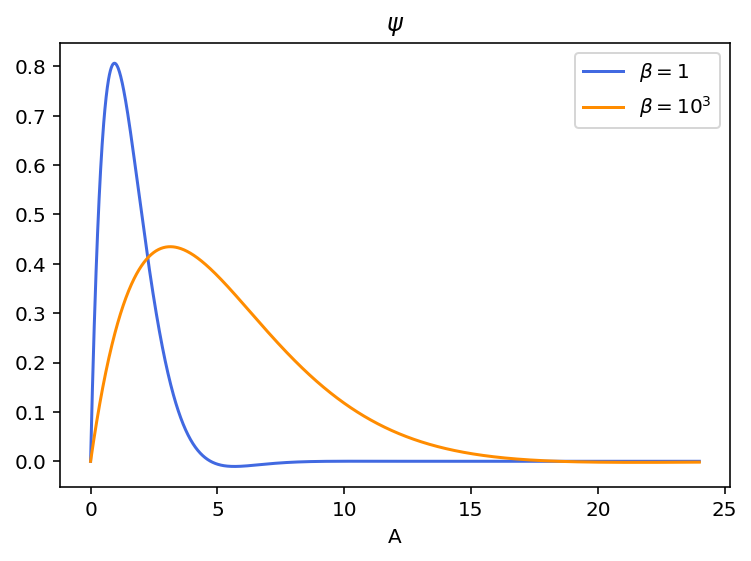}
	\caption{Normalized wave functions (\ref{psi}) with $\beta>0$ ($\Lambda=1=\kappa$). }
	\label{gig1}
\end{figure}

The resulting complex exponential wave functions (with simple ordering $n=0$) can be used to define a square-integrable wave function with hard-wall boundary condition $\Psi|_{A=0}=0$ given by 
\begin{align}\label{psi}
	\Psi(A)=\frac{2 \sqrt{\rho \cos \theta}}{\sin \theta} \, e^{-\rho A \cos \theta} \sin (\rho A \sin \theta ), 
\end{align}
and depicted in Fig. \ref{gig1}. It has a global maximum at $A_{\text{max}}=\theta/(\rho \sin \theta)$.

In contrast to $\hat{P}$, operator $\hat{P}^2$ admits self-adjoint extensions on the half-real line $A\ge 0$ (e.g., with Dirichlet boundary condition $\Psi(0)=0$) and represents a genuine observable,
\begin{align}\label{mean}
\braket{\hat{P}^2}=\hbar^2 \rho^2.
\end{align}

Finally, we describe an application of wave function (\ref{psi}) in connection with the problem of time \cite{kuchar}. Let's set $\Lambda=0$ and introduce a scalar field with a slowly varying potential (cf. Section \ref{sec6}). In a region of two-dimensional minisuperspace where we can neglect the matter kinetic term in the WDW equation, the wave function will be approximated by (\ref{psi}) with an effective cosmological constant $\Lambda(\phi)=\kappa^2 V(\phi)$, that is $\Psi(A,\phi)\approx \Psi(A)|_{\Lambda\to \Lambda(\phi)}$. Further, the maximum of $\Psi(A,\phi)$ will trace a trajectory in minisuperspace. Following \cite{PhysRevD.93.043505}, we define a time-dependent wave function as follows, 
\begin{align}
	\psi(A,t)=\left. \frac{\Psi(A,\phi)}{\sqrt{\int_0^{\infty} dA |\Psi(A,\phi)|^2}}\right|_{\phi=t}=\Psi(A,t)
\end{align}
since (\ref{psi}) is already normalized in $A$. The expectation value of $\hat{P}^2\propto \hat{H}^2$, thus becomes a function of time. Using definition (\ref{AP}) and the mean value (\ref{mean}), 
\begin{align}
	\braket{H^2}(t)&=\frac{\kappa^4 \hbar^2}{4} \rho^2|_{\Lambda(t)}.
\end{align}

\begin{figure}[h!]
	\centering
	\includegraphics[scale=0.54]{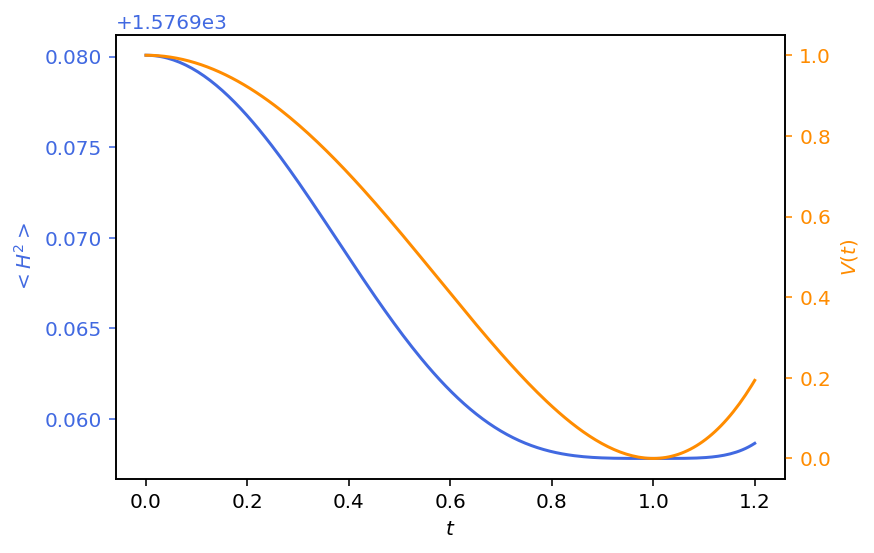}
	\caption{Mean value $\braket{H^2}(t)$ computed using $\psi(A,t)$ compared with the potential $V(t)=(\phi^2-1)^2$.}
	\label{gig14}
\end{figure}

Fig. \ref{gig14} shows $\braket{H}^2(t)$ for the scalar potential $V(\phi)=(\phi^2-\phi_0^2)^2$ (with $\phi_0=1 m_P$) and $\beta=10^{-9}$. The evolution displayed can be considered of inflationary type, in fact, integrating numerically $\int_0^1 \sqrt{\braket{H^2(t)}} dt$, we obtain around $40$ e-folds in the interval [0,1] which corresponds to the field traveling from the local maximum at $\phi=0$ to one of the minima at $\phi=1 m_P$. 

\section{Spatially closed universe}\label{sec5}
We can use the pair (\ref{AP}) with $k=1$, however, the inverse $a(A,P)$ satisfies an algebraic equation of degree four leading to a quite involved expression. An alternative is the following canonical pair,  
\begin{align}\label{newpairk}
	\begin{split}
		A'&=\frac{a^3 B (B^2+3)}{(B^2+1)^{3/2}} \left[1+48 \beta \kappa^2 \frac{(B^2+1)^2}{a^4}\right],\\
		P'&=-\frac{2 \sqrt{B^2+1}}{\kappa^2 a}
	\end{split}
\end{align}
Now $a^2(A',P')$ satisfies a cubic equation, thus not straightforward either, but at least we can factor out a term $a^3$ of the Hamiltonian constraint as with $k=0$. Thus, a WDW equation can be formulated that is identical to (\ref{wdw2}). The wave function  $\Psi(A')=\exp (i R A'/\hbar)$ with real $R$ are now associated to compact 3-geometry. Indeed, setting $P'=\frac{\p S_0}{\p A'}$ we get, in the gauge $N=1$, 
\begin{align}
R=-\frac{2 \sqrt{\at^2+1}}{\kappa^2 a}
\end{align}
which yields the $k=1$ solution in (\ref{classicsol}).

Some features of spatially closed models such as the Euclidean-Lorentzian boundary are, however, more transparent if we use the following pair
\begin{align}\label{dos}
\begin{split}
X&=\frac{3}{\kappa^2} a^2 \left[1-48 \beta \kappa^2 \frac{\left(B^2+1\right)^2}{a^4}\right],\\
\Pi&=-B.
\end{split}
\end{align}
In natural units ($c=1=\hbar$) $X$ is dimensionless since $\kappa^2=8 \pi l_P^2$ and the scale factor was redefined with length dimension in Section \ref{sec2}.
 
We restrict to negative $\beta$ since then $X$ is a positive variable. The inverse transformation reads 
\begin{align}\label{dosinverse}
\begin{split}
	a^2&=\frac{\kappa^2}{6} \left(X\pm \sqrt{X^2+1728 \beta \kappa^{-2} \left(1+\Pi^2\right)^2}\right),\\
	p_a&=\frac{6 a\, \Pi}{\kappa^2}+\frac{48 \beta}{a^3} \left(\frac{6}{5} \Pi^5+4 k \Pi^3+6 k^2 \Pi \right).
\end{split}
\end{align}
The branches of $a(X,\Pi)$ allow for the two solutions (\ref{alphaminus}). Indeed, on the $k=1$ solution (\ref{classicsol}), the square root in (\ref{dosinverse}) is proportional to the absolute value of $1+48 \beta \kappa^2 \alpha^4$ which, as seen in Fig. \ref{gig5}, can be either positive or negative. 

Transformations (\ref{dosinverse}) are canonical there is an $F(X,P)$ such that $p_a da-\Pi dX=dF$, namely, 
\begin{align}\label{canon}
	F&=\frac{2 \Pi^3 \left(5+3 \Pi^2\right)}{15 \left(1+\Pi^2\right)^2} \Big(\pm \sqrt{X^2+\tilde{\beta} (1+\Pi^2)^2}-X\Big) 
\end{align}
where $\tilde{\beta} \equiv 1728 \beta \kappa^{-2}$.

Expressing the Hamiltonian constraint (\ref{hamihhigher}) in terms of ($X,\Pi$), we get
\begin{align}\label{hamilk}
	\frac{H_0}{3 a}=-\frac{\Pi^2+1}{\kappa^2}-16 \beta \frac{\left(\Pi^2+1\right)^3}{a^4(X,\Pi)}+\frac{\Lambda a^2(X,\Pi)}{3 \kappa^2}, 
\end{align}
with $a^2(X,\Pi)$ given in (\ref{dosinverse}).

We will only consider WKB type wave functions $\Psi_{\text{WKB}}(X)=\exp[\frac{i}{\hbar} S(X)]$, on which we define $\widehat{a^2(X,\Pi)}$ as a series expansion in $\beta$, 
\begin{align}\label{series1}
\widehat{a^2} \propto X \Big[1 \pm \Big(1&+\frac{\tilde{\beta}}{2}  X^{-2} (1-\hbar^2 \p_X^2)^2 \nonumber \\
	& -\frac{\tilde{\beta}^2}{8}  \big(X^{-2} (1-\hbar^2 \p_X^2)^2\big)^2+...\Big)\Big] 
\end{align}
For a non-perturbative expression in $\beta$, after setting up the WKB expansion $S(X)=S_0(X)+\hbar S_1(X)+...$, we sum up the $\beta$-terms at each order in $\hbar$. Thus, for example, to zeroth-order in $\hbar$, $\widehat{a^2} \Psi=\Psi [X \pm (X^2+\tilde{\beta} (1+S_0'^2)^2)^{1/2}+\mathcal{O}(\hbar)]$,
where $S_0'\equiv \p S_0(X)/\p X$. In this way, we verify that the zeroth-order term $S_0(X)$ satisfies the Hamilton-Jacobi equation for (\ref{hamilk}), 
\begin{align}\label{hjkone}
1+S_0'^2&+\frac{\tilde{\beta} \big(1+S_0'^2\big)^3}{3 \Big(X\pm \sqrt{X^2+\tilde{\beta} (1+S_0'^2)^2}\Big)^2} \nonumber \\ 
& -\frac{\Lambda \kappa^2}{18} \Big(X\pm \sqrt{X^2+\tilde{\beta} (1+S_0'^2)^2}\Big)=0
\end{align}

As is well known, $S_0(X)$ is related to the action. More precisely, using (\ref{canon}),  
\begin{align}\label{so}
    S_0(X)&=\int \left(\Pi \dot{X}-H(X,\Pi)\right) dt \nonumber \\
    &=\int \left[\at p_a-\tilde{H}(a,p_a)\right] dt-F.
\end{align}
Evaluating (\ref{so}) at the $k=1$ solution (\ref{classicsol}) yields the zeroth-order WKB term,
\begin{align}\label{zeroth}
S_0(X)=\pm \frac{2 \bar{X}}{3} \left(\frac{X}{\bar{X}}-1 \right)^{3/2}+\text{constant},
\end{align}
where
\begin{align}\label{xmin}
	\bar{X}&=\frac{3 (1-48 \beta \kappa^2 \alpha^4)}{\kappa^2 \alpha^2}
\end{align}
and $\alpha$ is either of the roots in (\ref{alphaminus}).

The characteristic scale  $\bar{X}$ is defined by the vanishing of $\Pi$. In fact, setting $\Pi=S'_0=0$ in (\ref{hjkone}) yields an algebraic equation for $\bar{X}$ with solution (\ref{xmin}).  

Now, $X<\bar{X}$ is a classically forbidden region, nonetheless, we can analytically continue (\ref{zeroth}), 
\begin{align}\label{cont}
S_{0}(X)=\pm i \frac{2 \bar{X}}{3} \left(1-\frac{X}{\bar{X}} \right)^{3/2}+\text{constant},
\end{align}
which yields exponentially decaying/growing wave functions. 

The 4-geometry associated to (\ref{cont}) is the 4-sphere of radius $\alpha^{-1}$. To see this, we identify $S_0'$ from (\ref{cont}) with the momentum (\ref{dos}), that is, $\frac{\p S_0}{\p X}=-B=-\frac{\at}{N}$. Then, setting $N=i$, which corresponds to proper Euclidean time $\tau$, we get
\begin{align}
\frac{da}{d\tau}=\pm \Big[1-\frac{\alpha^2 a^2 \big(1-48 \beta \kappa^2 a^{-4} \big(1-(\frac{da}{d\tau})^2\big)^2 \big)}{1-48 \beta \kappa^2 \alpha^4} \Big]^{1/2}
\end{align}
whose integration yields $a(\tau)=\frac{1}{\alpha} \cos(\alpha (\tau-\tau_0))$.  Therefore, $\bar{X}$ marks the transition from Euclidean to Lorentzian geometry. 

Now, we continue with the first-order WKB term. Defining left-hand side of the H-J equation  (\ref{hjkone}) as $T(S_0',X)$, the equation satisfied by $S_1(X)$ is
\begin{align}\label{first}
S_1' \frac{\p T}{\p S_0'}=\frac{i}{2} S_0^{''} \frac{\p^2 T}{\p S_0'^2}.
\end{align}
Setting $T=0$ after taking its derivatives and using the zeroth-order term (\ref{cont}), we obtain
\begin{align}
	S_1(X)=\frac{i}{4} \ln  \left| \frac{X}{\bar{X}}-1\right| 
\end{align}

Therefore, the first-order WKB wave functions are
\begin{subequations}\label{wkb}
\begin{align}
&\Psi_{\text{WKB}(1)}=\frac{\exp \big[c\pm i \frac{2 \bar{X}}{3 \hbar} \left(\frac{X}{\bar{X}}-1 \right)^{3/2}\big]}{\big(X/\bar{X}-1\big)^{1/4}}, && \bar{X}<X \\
&\Psi_{\text{WKB}(1)}=\frac{\exp \big[d\pm \frac{2 \bar{X}}{3 \hbar} \left(1-\frac{X}{\bar{X}} \right)^{3/2}\big]}{\big(1-X/\bar{X}\big)^{1/4}}, && X<\bar{X} 
\end{align}
\end{subequations}
where $c, d$ are normalization constants. (\ref{wkb}) resembles ordinary FRW wave functions \cite{wiltshire2003introductionquantumcosmology}, except that $X$ and $\bar{X}$ depend, in a rather involved way, on the coupling constant $\beta$, whereas with the $\beta=0$, $X\propto  a^2$ and $\bar{X}=9/\kappa^2 \Lambda$.

Since the WKB approximation does not hold for $X\approx \bar{X}$, the usual approach is to find a solution of the WDW equation valid to all orders of $\hbar$ in a small neighborhood of $\bar{X}$ \cite{liboff}. Since we have a non-standard Hamiltonian (\ref{hamilk}), we first expand it around the phase space point $(X=\bar{X},\Pi=0)$, retaining only linear terms in $X-\bar{X}$ and quadratic in $\Pi$. The WDW equation of the linearized Hamiltonian is 
\begin{align}\label{airy}
\frac{d^2\Psi}{dy^2}-y \Psi=0 
\end{align}
where $y=-(\hbar^2 \bar{X})^{-1/3} (X-\bar{X})$.

\begin{figure}[h!]
	\centering
	\includegraphics[scale=0.6]{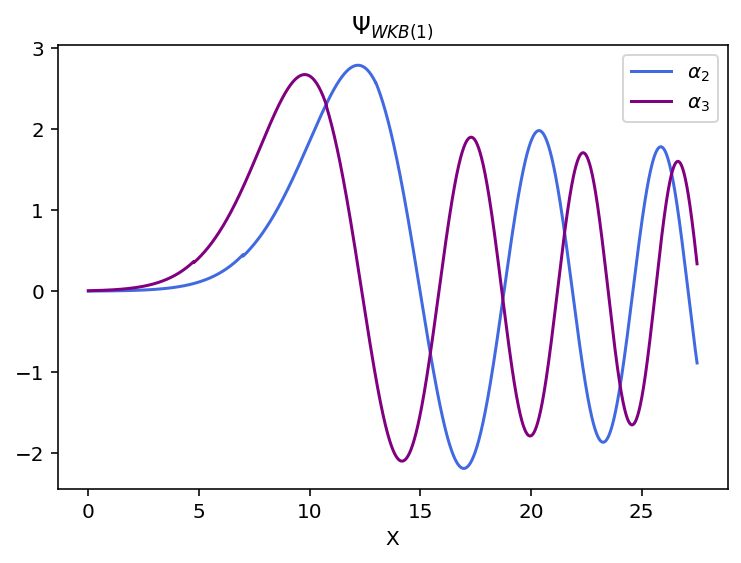}
	\caption{Real WKB solution (\ref{wkbpatch}) connected by the (linearized) solution $2\sqrt{\pi} \bar{X}^{1/6} Ai (y)$ around $\bar{X}_2\approx 9.99$ and $\bar{X}_3=7.76$.}
	\label{gig11}
\end{figure}

Two independent solutions of (\ref{airy}) are the Airy functions $Ai(y), Bi(y)$. The solution that goes to zero as $|y| \to \infty$ is the Airy function $Ai(y)$. Comparing the asymptotic forms of $Ai(y)$ for large $|y|$ \cite{liboff} with (\ref{wkb1}) and (\ref{wkb2}) we obtain the particular solution  
\begin{subequations}\label{wkbpatch}
	\begin{align}
		&\Psi_{\text{WKB}(1)}=\frac{2 \sin  \big[\frac{2 \bar{X}}{3 \hbar} \left(\frac{X}{\bar{X}}-1 \right)^{3/2}+\frac{\pi}{4}\big]}{\big(X/\bar{X}-1\big)^{1/4}}, && \bar{X}<X \\
		&\Psi_{\text{WKB}(1)}=\frac{\exp \big[-\frac{2 \bar{X}}{3 \hbar} \left(1-\frac{X}{\bar{X}} \right)^{3/2}\big]}{\big(1-X/\bar{X}\big)^{1/4}}, && X<\bar{X} 
	\end{align}
\end{subequations}
which is depicted in Fig. \ref{gig11}.

\section{Scalar field}\label{sec6}
Although CECG was introduced as a potential inflationary model in the presence of ordinary matter, it is interesting to see how scalar-field driven inflation is affected by cubic curvature terms.
Thus we consider a homogeneous scalar field 
\begin{align}\label{scalarfield}
L^\phi=N a^3 \Big[\frac{\dot{\phi}^2}{2 N^2}-V(\phi)\Big] 
\end{align}
with scalar potential 
\begin{align}\label{star}
	V(\phi)=\frac{3 M^2}{4 \kappa^2} \big[1-e^{-\sqrt{2/3} \kappa \phi} \big]^2,
\end{align}
($M\approx 10^{13} GeV\approx 10^{-6} m_P$) which is well-known from Starobinsky inflation in its scalar-tensor version \cite{defelice2010,ketov2025legacystarobinskyinflation}.

The Friedmann equations can be obtained replacing $\Lambda$ in (\ref{friedcanon}) with the energy density $\kappa^2 \rho_\phi \equiv \kappa^2 (\frac{1}{2} \dot{\phi}^2+V(\phi))$ and replacing $\Lambda$ in (\ref{acceleration}) with the pressure $-\kappa^2 p_\phi \equiv -\kappa^2 (\frac{1}{2} \dot{\phi}^2-V(\phi))$. Additionally, we have now the Klein-Gordon equation $\ddot{\phi}+3 H \dot{\phi}+\frac{dV}{d\phi}=0$, where $H=\at/a$ is the Hubble factor.

\begin{figure}[h!]
	\centering
	\includegraphics[scale=0.6]{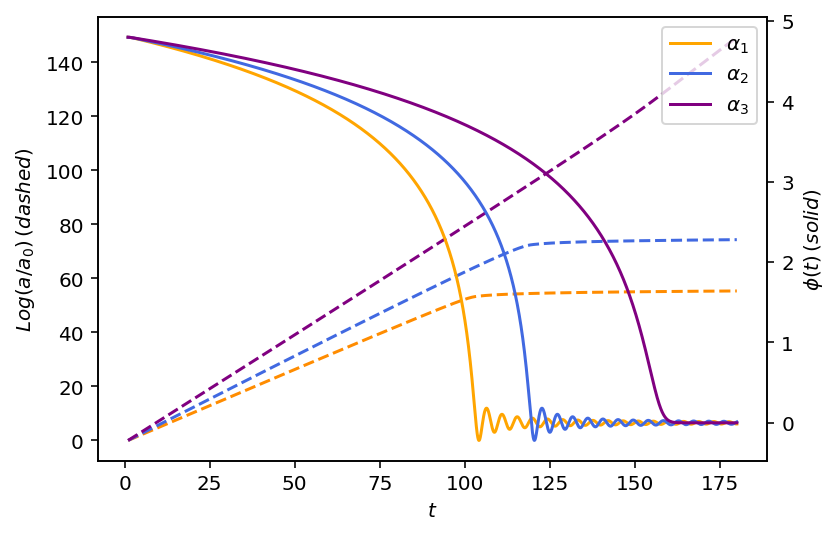}
	\caption{Inflationary evolution with $\beta=0.1$ and $\beta=-0.075$ for scalar potential (\ref{star}). }
	\label{gig12}
\end{figure}

Some numerical solutions of the equations of motion are presented in Fig. \ref{gig12}. We fix the initial values of $(a, \phi, \dot{\phi})$, and use the Hamiltonian constrain to get the initial Hubble factor. For this, we replace $\Lambda$ by $\kappa^2 (\frac{1}{2} \dot{\phi}^2+V(\phi))$ on the right-hand side (\ref{alphaplus}) (or (\ref{alphaminus}), which give us $\alpha^2=H^2+\frac{k}{a^2}$.  Thus, for negative $\beta$, there are two possible initial values of $H$ leading to the solutions labeled $\alpha_2, \alpha_3$ in Fig. \ref{gig12}. The solutions $\alpha_1$ (for positive $\beta$) and $\alpha_2$ are qualitatively similar to the ordinary case $\beta=0$, with a slightly different number of e-folds. In this case, the scalar field undergoes weak damping at the end of inflation. With the solution $\alpha_3$, however, the scalar field undergoes a strong damping that quickly drains the total field's energy. In this case, the scale factor evolves according to (\ref{vaccum}) with $H^2=(-16 \beta \kappa^2)^{-1/2}$, and does not exit inflation. 

\subsection{Quantization}
Using the canonical pair (\ref{dos}), the Hamiltonian of FRW CECG plus scalar field is proportional to
\begin{align}\label{hamilphi}
\frac{\Pi^2+1}{\kappa^2}+16 \beta \frac{(\Pi^2+1)^3}{a^4}-\frac{\frac{1}{2} p_\phi^2+a^6 V(\phi)}{3 a^4}=0
\end{align}
where $a=a(X,\Pi)$ given in (\ref{dosinverse}).

We will consider WKB type wave functions in the region of minisuperspace ($X,\phi$) where the potential (\ref{star}) varies slowly, thus allowing us to neglect the scalar field kinetic term $p_\phi^2$. Following the analysis of the quantum Starobinsky model \cite{mijic}, to zeroth-order in $\exp(-\sqrt{2/3} \kappa \phi)$, the potential (\ref{star}) is constant and the WDW equation corresponding to (\ref{hamilphi}) is separable. Writing $\Psi(X,\Phi)=\mathcal{A}(X) \Phi(\phi)$, the matter wave function satisfies $(\hat{p}_\phi^2-\xi) \Phi(\phi)=0$, where $\xi$ is the separation constant. Then, by choosing $\xi=0$ and $\Phi(\phi)=\text{constant}$, we are effectively discarding any contribution from $\hat{p}_\phi^2$. On the other hand, the gravity component of the wave function is $\mathcal{A}(X)=\Psi_{\text{WKB}(1)}(X)$ in (\ref{wkb}) with $\Lambda=\kappa^2 V \approx \frac{3}{4} M^2$.

\begin{figure}[h!]
	\centering
	\includegraphics[scale=0.6]{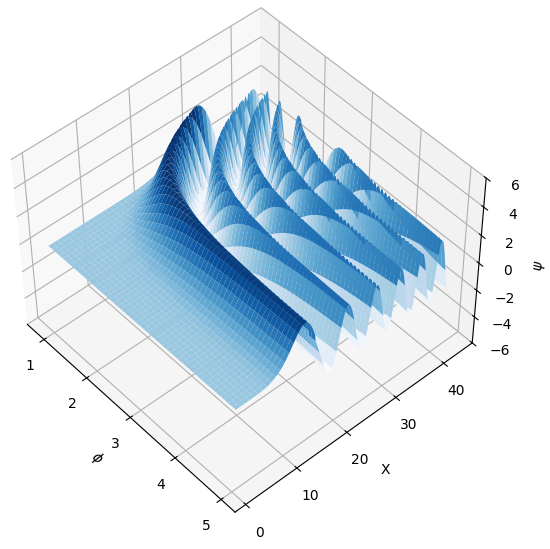}
	\caption{WKB wave function on the two-dimensional minisuperspace ($\phi,X$).}
	\label{gig4}
\end{figure}

To take into account the scalar field $\phi$, we substitute the potential to first-order in $e^{-\sqrt{2/3} \kappa \phi}$. The WDW equation is no longer separable, but the contribution from $\hat{p}_\phi^2$ term is of order $X^{-2} \exp(-\sqrt{2/3} \kappa \phi)$. Thus, one restricts to a region of superspace where this term can be neglected ($X\gg 1$) and the WKB solutions still have the form (\ref{wkb}), namely, 
\begin{align}\label{wkb1} 
\Psi_{\text{WKB}}(X,\phi)=\frac{\exp \big[\pm i \frac{2 \bar{X} (\phi)}{3 \hbar} \big(\frac{X}{\bar{X}(\phi)}-1\big)^{3/2}+c(\phi) \big]}{\big(\frac{X}{\bar{X}(\phi)}-1\big)^{1/4}}, 
\end{align}
for  $X>\bar{X}(\phi)$, and 
\begin{align}\label{wkb2}
	\Psi_{\text{WKB}}(X,\phi)=\frac{\exp \big[\pm \frac{2 \bar{X} (\phi)}{3 \hbar} \big(\big(1-\frac{X}{\bar{X}(\phi)}\big)^{3/2}+d(\phi)\big]}{\big(1-\frac{X}{\bar{X}(\phi)}\big)^{1/4}},
\end{align}
for $X<\bar{X} (\phi)$. We can use them to construct wave functions such as the one depicted in Fig. \ref{gig4}, which follows by changing $\Lambda$ in (\ref{wkbpatch}) by $\kappa^2 V(\phi)$.

\begin{figure}[h!]
	\centering
	\includegraphics[scale=0.54]{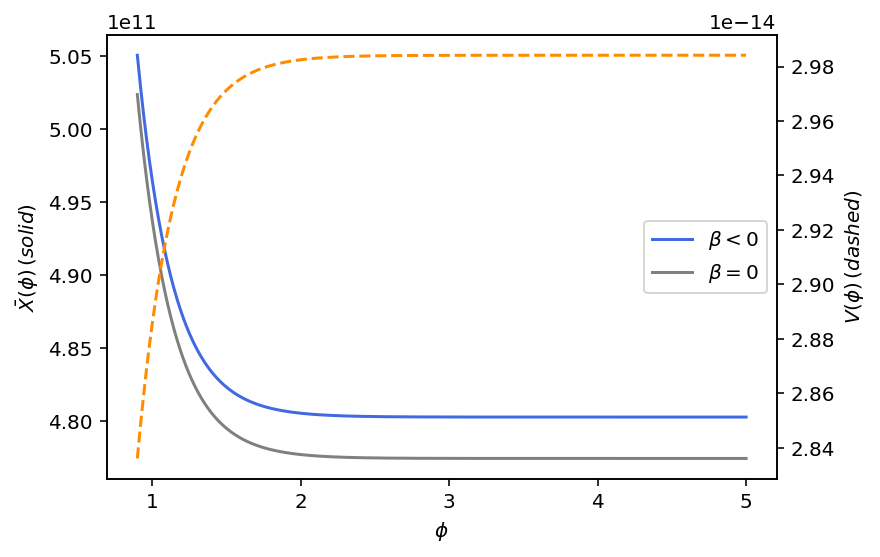}
	\caption{$\bar{X}(\phi)$ compared to the scalar potential. }
	\label{gig3}
\end{figure}

The function $\bar{X}(\phi)$, depicted in Fig. \ref{gig3}, is also defined by equations (\ref{xmin})-(\ref{alphaminus}), upon making the following replacement, $\Lambda\to \kappa^2 V(\phi)=\frac{3}{4}  M^2 [1-2 \exp(-\sqrt{2/3} \kappa \phi)]$.

The integration constants in (\ref{wkb}) are now the functions of $\phi$. Other choices of $c(\phi)$ and $d(\phi)$ depend on the boundary conditions imposed on the wave function. For instance, the Hartle-Hawking no-boundary proposal requires $\Psi$ independent of $\phi$ as the scale factor goes to zero \cite{moniz},  or in this case, as $X\to 0$. This yields $d(\phi)=\mp \frac{2 \bar{X} (\phi)}{3 \hbar}$, and similarly for $c(\phi)$. A rigorous derivation, however, requires evaluating the path integral and will be addressed in subsequent investigations. 

The curve $\bar{X}(\phi)$, roughly the projection on the plane $(\phi,X)$ of the first crest in Fig. \ref{gig4}, separates regions of quantum-classical behavior of the universe. In fact, the phase of oscillating WKB solutions, $S_0 (X,\phi)=\frac{2}{3} \bar{X} (\phi) (X/ \bar{X}(\phi)-1 )^{3/2}$, predicts a strong correlation between $X, \phi$ and their momenta, according to the relations  $\Pi=\frac{\p S_0}{\p X}$ and $p_\phi=\frac{\p S_0}{\p \phi}$, which translate into 
\begin{align}
\begin{split}
	\at &=\pm \left(X/\bar{X}(\phi)-1\right)^{1/2}, \\
	a^3 \dot{\phi}&	\approx -\left(X/\bar{X}-1\right)^{1/2} \frac{d \bar{X}}{d \phi} \approx 0
\end{split}
\end{align}
using $\frac{d\bar{X}}{d\phi} \propto \frac{d V}{d \phi}$. Using (\ref{xmin}), the first equation yields $\alpha^2(\phi)=\frac{\at^2+1}{a^2}$, which implies equation (\ref{cubicw}) with $\Lambda \approx \kappa^2 V(\phi)$.

\section{Conclusions}\label{sec7}
We studied the Wheeler-DeWitt quantization of the FRW minisuperspace model corresponding to Cosmological Einsteinian Cubic Gravity. With only a positive cosmological constant, the classical dynamics is very similar to that of ordinary FRW models, with the cubic curvature terms inducing an effective cosmological constant depending on the coupling constant $\beta$. A natural question thus arises, how the higher-curvature terms affect the quantization method and the outcomes, in particular, the signatures of CECG on the wave function of the universe.

We find that already the canonical formulation of FRW CECG is not as straightforward as that of the ordinary FRW model. This comes about because the Lagrangian has a non-standard kinetic term that induces a nontrivial deformation of the ordinary symplectic structure: $\{a,p\}=1 \to \{a, p\}\propto 1/[1+48 \beta \kappa^2 p^4/a^8]$, where $p\propto a \at$. The actual variable, canonically conjugated to $a$, is a quintic polynomial in $\at$. Since there is no general solution to the quintic equation, we decided to implement different canonical transformations ($A,P$). In this way, we shift the problem of solving for $\at(a,p_a)$, for that of finding the inverse expression $a(A,P)$, which involves an algebraic equation of lower degree. The cost of using alternative canonical pairs is a non-polynomial Hamiltonian constraint, which renders the Wheeler-DeWitt quantization more challenging.

For instance, the Hamiltonian constraint $H_0(A,P)$ of the spatially flat ($k=0$) model carries the overall factor $D^{-1}=(1+\tilde{\beta} P^4)^{-1}$ which, upon quantization, becomes a non-local operator related to integration. Our approach was to choose a convenient ordering that in practice amounts to multiply the classical Hamiltonian by $D$, thus we quantize $a^{-3} H_0$ instead of $H_0$. Further, in the position representation, the quantum constraint $\widehat{a^{-3} H_0} \Psi=0$ translates into a sixth-order differential equation. Therefore, although CECG has no more degrees of freedom than the scale factor, we get a ``higher-derivative WDW equation". Thus, in  this case we obtained six independent solutions instead of two of ordinary FRW models. For example, with positive $\beta$, we get two purely imaginary exponential wave functions associated with classical solutions of the Friedmann equations, and four complex exponential wave functions associated with classically forbidden complex 4-geometries.

For the quantization of the spatially closed model, we use another canonical pair $(X,\Pi)$. We obtain WKB-type wave functions. In this case, we defined $a(X,\Pi)$ as a series in $\beta$. After setting up the WKB expansion, we sum up the $\beta$ terms at each order in $\hbar$ to obtain a non-perturbative expression. The wave functions obtained in this way depend on a certain scale $\bar{X}$ defined by the classical Hamiltonian constraint with vanishing momentum. While it resembles the usual WKB wave functions, the new configuration space variable $X \propto a^2(1-\beta P^4)$, represents a different physical quantity when $\beta \ne 0$. Finally, we consider inflation driven by a homogeneous scalar field. We obtain WKB type solutions neglecting the scalar field kinetic term and replacing $\Lambda$ with a slowly varying potential. In this case $\bar{X}(\phi)$ traces a trajectory in the two-dimensional minisuperspace separating the regions of classical/quantum behavior.

Several aspects were left out in this work. For instance, the path integral formulation of the wave function and the implementation of boundary conditions such as the Hartle-Hawking no-boundary or Vilenkin's quantum tunneling in the presence of cubic curvature terms. A rigorous treatment of the inverse operator and its connection with the order of WDW equation and the boundary conditions. An analysis of the deformed commutation relation in the context of Generalized Uncertainty Relations \cite{PhysRevD.97.126010,PhysRevD.94.123505}. The quantization of the actual geometric inflation models with curvature terms of all-orders will be subject of upcoming work.

\ \ \\

\centerline{\bf Acknowledgements}

N.E. Martínez-Pérez thanks SECIHTI for financial support.

\bibliography{bibliography}

\end{document}